\documentclass[onecolumn]{aa}
\usepackage{graphicx}
\usepackage{natbib}
\bibpunct{(}{)}{;}{A}{}{,}
\begin{document}
   \title{New candidate GHz Peaked Spectrum 
          and Compact Steep Spectrum Sources}

   \titlerunning{New candidate GPS and CSS Sources}

   \author{P.G. Edwards \inst{1} \and S.J. Tingay \inst{2}}

   \offprints{P.G. Edwards}

   \institute{Institute of Space and Astronautical Science, 
              Japan Aerospace Exploration Agency, \\
              Yoshinodai 3-1-1, Sagamihara, Kanagawa 229-8510, Japan \\
              \email{pge@vsop.isas.jaxa.jp}
         \and
              Centre for Astrophysics and Supercomputing, 
              Swinburne University of Technology, \\ P.O. Box 218, Hawthorn,  
              Vic. 3122, Australia\\
              \email{stingay@astro.swin.edu.au}
             }

   \date{Accepted 21 May 2004}

   \abstract{
Data from a recent Australia Telescope Compact Array (ATCA)
program of multi-frequency, multi-epoch
monitoring of 202 active galactic nuclei with declinations $< +10^\circ$
have been 
searched for GHz Peaked Spectrum (GPS) 
and Compact Steep Spectrum (CSS) radio sources. 
Supplementary data at higher and lower frequencies, where available, 
have been used to further examine the spectral properties of previously
reported and new candidate GPS and CSS sources.
The ATCA monitoring program has allowed the variability and polarization
properties of sources previously reported as GPS and CSS sources, and
the majority of new GPS and CSS candidates, to be investigated, confirming
that these are useful diagnostics in discriminating genuine
GPS and CSS sources from variable sources that display similar spectra
only temporarily. GPS sources are confirmed to be generally more
compact, and less polarized, than CSS sources,
although CSS sources show evidence for being somewhat
less variable than GPS sources at 1.4 and 2.5\,GHz.
In addition, the widths of GPS spectra are examined,
and a significant difference is found in the GPS sample of
Snellen et al.\ (2000) between sources with 
compact double (CD) or compact symmetric object (CSO) 
morphologies and sources with
other morphologies, in that CD and CSO sources have generally
narrower spectra. Possible reasons for this difference are considered.

   }

   \maketitle
%
%________________________________________________________________

\section{Introduction}

The classifications of GHz Peaked Spectrum (GPS) and Compact Steep
Spectrum (CSS) radio sources are based on their total flux density
spectra.  The conventional, though somewhat arbitrary, classification
scheme assigns the GPS label to sources with spectra which peak above
500\,MHz, and the CSS label to sources peaking at lower frequencies
(if a turnover can be detected at all).  In addition to the spectral
properties, morphological criteria are required for bona fide members
of these classes, with GPS sources generally 
less than 1\,kpc in projected linear size and CSS sources less than
20\,kpc.  An observed anti-correlation between turnover frequency and
projected linear size \citep{fan90,ode97} has led to a model in which
GPS sources evolve into CSS sources, and possibly then into FR-I
and/or FR-II sources \citep{sne03}.

Studies of the motion of the components in 
the Compact Symmetric Object (CSO) sub-class of 
GPS/CSS sources indicate that they 
have become active in the last $\sim 10^3$ years \citep{con02}.
In some cases this appears to be the first instance of activity;
in other sources there is evidence that the sources have been
re-activated after a period of quiescence
\citep[e.g.,][]{tin03a}. There is substantial
evidence that mergers or other disruptions to the galaxy core
have produced this activity.  
GPS radio sources are identified with both quasars and galaxies,
and while GPS quasars share many of
the radio properties of GPS galaxies, 
it has been argued that they are probably not a single
class of object unified by orientation \citep{sne99}.

Early interest in GPS sources stemmed from the fact that 
a significant fraction were identified with high redshift sources. 
This resulted in searches for GPS sources deliberately
excluding sources identified with galaxies 
\citep{gop83,spo85}.
More recently, the focus of attention has shifted to source evolution,
as the youth of (at least the CSO sub-class of) GPS sources 
provides the opportunity to study radio sources in early stage
of development.
GPS galaxies also have the advantage that high linear resolution 
observations can be obtained of the central regions 
\citep[e.g.,][]{sne03,tin03a}.

The generally accepted criteria for identifying GPS sources are:
a spectral turnover at GHz frequencies, with 
a spectral index ($S \propto \nu^{+\alpha}$) 
at low frequencies of $\alpha > +0.5$ or
a spectral index at high frequencies of $\alpha < -0.5$, and 
a spectral curvature of at least 0.6 (see \S2).
Sources with spectra that turn up at lower or higher frequencies were
initially excluded from GPS catalogues 
\citep{gop83}, although it is becoming
increasingly clear that this requirement may be too restrictive.
In particular, the spectra of re-activated sources may have a turn up at
lower frequencies corresponding to extended emission
resulting from a previous active period.
In addition to these criteria, GPS generally display at least some of
the following properties:
compact structure (confined to the central 1\,kpc),
usually resolved into two or three components by VLBI,
low fractional radio polarization,
low variability, and sub-luminal component motions.
Compact Steep Spectrum sources share many of properties of GPS sources,
with the main differences being, as mentioned above, that 
the spectral turnover, if present,
occurs at lower frequencies, 
that CSS sources have higher fractional polarizations,
and that the linear size is somewhat
larger \citep[see, e.g.,][]{ode98}.

In this paper we use data from a multi-frequency, multi-epoch
survey \citep{tin03b},
undertaken with the Australia Telescope Compact Array (ATCA),
to search for new GPS and CSS candidates. 
The survey, which was made in conjunction with the VSOP Survey Program
\citep{hir00},
originally contained 202 sources, however 17 of these were dropped
after the first few epochs as it was clear they did not meet the
criteria for inclusion in the VSOP Survey Program. The remaining 185
sources were observed at up to 16 epochs between 1996 and 2000 at  
1.384, 2.496, 4.800 and 8.640\,GHz,
with both total and polarized flux densities being measured.
The multi-epoch observations allowed, for each source, an estimate
of the variability index, $m$,
defined as the r.m.s.\ variation
from the mean flux density divided by the mean flux density. 
Sources were not imaged, but a flag was assigned based on the degree of 
compactness: 
`c' for compact sources, `e', extended, and `l', showing some sign of
extension at low frequencies.
The approximate angular resolution on the ATCA 6\,km baseline is
7$''$ at 1.4\,GHz, 4.5$''$ at 2.5\,GHz, 2$''$ at 4.8\,GHz, 
and 1$''$ at 8.6\,GHz.

In the Tables in this paper 
we reproduce the values from \cite{tin03b} for
the source characteristics at 4.8\,GHz,
as this frequency band is commonly used to characterise
sources \citep[see, e.g.,][]{ode98},
and as it is close to the peak frequency for many of the GPS
sources we consider.
The mean variability indices at 4.8\,GHz for the 185 sources
ranged between 0.01 and 0.43, 
with a median value of 0.08.
The highest observed linear polarization was 8.09\% at 4.8\,GHz,
with a median level of 2.21\%.
Twenty-four sources (13\% of the 185 sources) had polarization levels
below the reliably measurable level of 0.5\%. 

In addition to the ATCA data, we have used data from 
RATAN-600 observations at
0.96, 2.30, 3.90, 7.70, 11.2, and 21.65\,GHz
\citep{kov99},
VLA calibrator observations, particularly post-1998 43\,GHz data,
(http://www.aoc.nrao.edu/$\sim$gtaylor/calib.html),
ATCA calibrator observations, particularly at 22\,GHz 
(http://www.narrabri.atnf.csiro.au/calibrators/),
843\,MHz SUMSS survey data  \citep{mau03},
archival 408\,MHz data from the MRC catalog  \citep{lar81},
archival 365\,MHz data from Texas catalog  \citep{dou96},
and archival 160\,MHz data from the Culgoora catalog  \citep{sle95}.
With the exception of the last three data sets, 
it has been attempted to use contemporaneous data as much as possible.
The phenomenon of low frequency variability notwithstanding,
use of the archival data at frequencies below 843\,MHz is acceptable
as sources are expected to show less intrinsic variability at
these frequencies.
In several instances, other data have been used,
and, in addition, a handful of 22\,GHz observations were made with the ATCA 
in April 2003 specifically for this study.

The ATCA monitoring was undertaken for sources south of a declination of
$+10^\circ$, and so this study supplements the work of
\cite{kin97} and \cite{sne02} in redressing
the under-representation of southern sources in GPS and CSS catalogs
\citep[see, e.g.,][]{ode98}.

In order to compare the characteristics of new candidate GPS and
CSS sources with established members of these classes, we first
examine the properties of previously reported sources before
presenting the same properties for the new candidates.
In \S2 we study 20 previously reported GPS sources, and in \S3
we present data for 8 new GPS candidates. In \S\S4 and 5 we examine
8 previously reported, and 12 new candidate, CSS sources. In \S6 we
present the properties of 
a number of other sources of interest, which we find do not 
meet our criteria for inclusion as GPS or CSS candidates.
After tabulating the properties of sources in each section,
short comments are given for individual sources. These 
draw attention to particular properties of the source revealed
by the ATCA monitoring, and are accompanied by illustrative,
rather than exhaustive, notes from the literature.

%__________________________________________________________________

\section{Previously reported GPS sources}

\begin{figure*}
   \centering
   \includegraphics{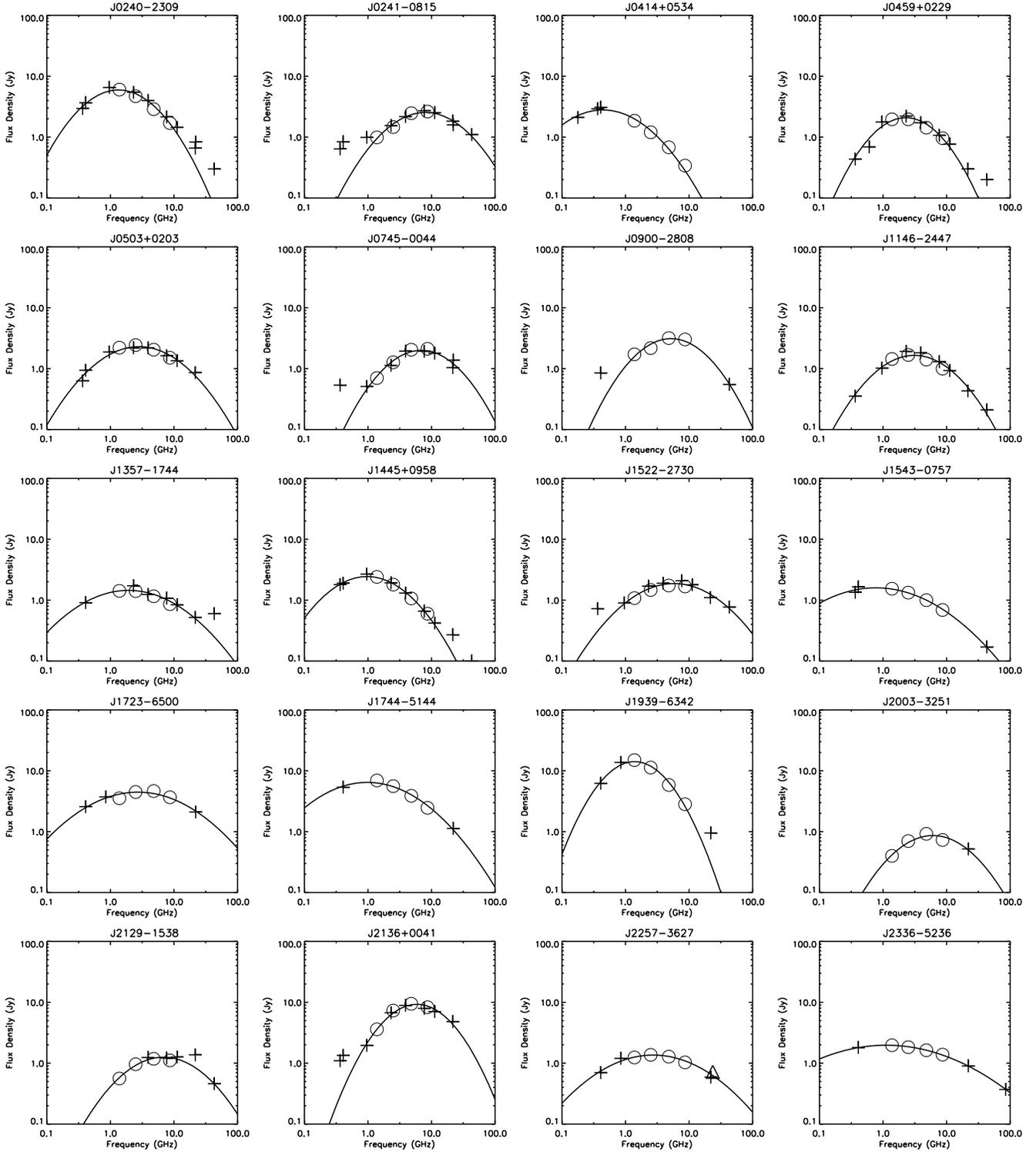}
   \caption{Spectra of previously identified GPS sources.
Circles denote data from \cite{tin03b} and plus symbols
denote data from the other sources described in \S1. 
The triangles denote 22\,GHz observations made with the ATCA in April 2003
as part of this study. Any additional
sources are listed in the Notes on individual sources.
The fit to the spectra is described in \S2 and
fit parameters are given in Table~\ref{tab1}.
}
              \label{Fig1}%
    \end{figure*}

The properties of GPS sources have been comprehensively reviewed by 
\cite{ode98}. In Table~\ref{tab1} we list those known GPS sources which were
included in the ATCA program, and plot their spectra in Figure~\ref{Fig1}.
The references given in Table~\ref{tab1} are
to compilations of GPS sources which include
the source in question, and generally, but not always, include
the original report of discovery of the GPS-like nature of the source.

We also list in Table~\ref{tab1} the source classification and redshift,
and the mean flux density, $S$, mean fractional
polarization, $p$, and variability index, $m$, all at 4.8\,GHz, from the ATCA
monitoring data.
Where possible, we also give for comparison, 
in brackets, the polarized flux measured
with the VLA at 4.9\,GHz \citep{per82}.
The spectral index at high frequency, $\alpha_{hi}$, is generally determined
from the data at the highest 
frequencies plotted for that source in Figures~\ref{Fig1} and~\ref{Fig2}.
The spectral curvature, $\Delta\alpha$, is the absolute difference
between the high frequency and the low frequency spectral indices,
following \cite{dev97}.
The width is the FWHM of the fitted spectrum in decades of frequency,
following \cite{ode91}.
The flag is that from the ATCA data described in \S1.

De Vries et al.\ (1997) have examined the spectra of GPS sources
and derived an average spectrum with a optically thick spectral
index of 0.5 below the peak, and a broken power-law spectrum
above the peak changing from $-$0.4 to $-$0.7 at twice the peak
frequency. However, here, for simplicity we have fitted a spectrum 
of the type
$$\log S = a (\log\nu)^2 + b \log\nu + c$$
to the data for GPS sources
\citep[as also used, e.g., by][]{pan98}. 
No physical meaning is ascribed to this fit: it is adopted only to
estimate the peak frequency and the width of the spectrum in a
consistent manner for all sources. It is clear from Figure~\ref{Fig1} that a
spectrum of this form does not provide an accurate fit at frequencies
far from the peak, and in some cases there are points which deviate
from any simple spectral fit. As we are interested in the shape of the
spectrum near the peak we have simply excluded these points in the
spectral fit: the points neglected in this way are easily recognized
in Figures~\ref{Fig1} and~\ref{Fig2}.

It is notable that the variability index, $m_{4.8}$, is less than
or equal to the median value (0.08) for 18 of the 20 sources.
In addition, the fractional polarization, $p_{4.8}$, is less than
median value for 17 of the 20 sources. Indeed, half of the 24 sources
for which only an upper limit could be set to the fractional
polarization are found in this Table. Furthermore, all seven
sources identified with galaxies have $p_{4.8} < 0.5$.
The one source that exceeds
the median in both cases is the BL~Lac object 
J1522$-$2730, which we consider in more detail below.
Comparison of the structure flags is also interesting:
of the 202 sources in the sample of \cite{tin03b},
about 50\% had a `c' flag, 30\% an `e' flag, and 20\%
an `l' flag. 
In contrast, of the 20 sources in Table~\ref{tab1},
14 have `c' flags, one has an `e' flag, and 5 have `l' flags.
These results lend confidence to the use of these parameters
in selecting GPS sources.

%                                     Two column figure (place early!)
%______________________________________________ Gamma_1 (lg rho, lg e)

%__________________________________________________ One column table
\begin{table*}
\caption[]{Properties of previously identified GPS sources and candidates.
Type Q denotes quasar; G, galaxy; GL, gravitational lens; and BL, BL~Lac
object. Other parameters are described in \S2.
The percentage polarized flux densities given in brackets are those
measured at 4.9\,GHz by \cite{per82}.
}
\label{tab1}
\begin{tabular}{l l l c c c c rl c c c c c}
\hline
\noalign{\smallskip}
Source & B1950 or & Refs\,$^{\mathrm{a}}$ & Type & $z$ & $S_{4.8}$ & $m_{4.8}$ & \multicolumn{2}{c}{$p_{4.8}$} & $\nu_{pk}$& $\alpha_{hi}$ & $\Delta\alpha $ &width & flag \\
       & Other    &      &      &       & (Jy)  &       & \multicolumn{2}{c}{(\%)}  & (GHz)     &               &                 &      &  \\
\noalign{\smallskip}
\hline
\noalign{\smallskip}
J0240$-$2309& 0237$-$233&K97V97O98 &Q & 2.223& 2.88& 0.03&     3.47&(4.2)&1.3 &$-$1.1&1.6 &1.2& c \\
J0241$-$0815&  NGC1052  &V97       &G & 0.004& 2.47& 0.09& $<$0.5\,~~&(0.3)&7.5 &$-$0.6&1.2 &1.4& c \\
J0414$+$0534& 4C+05.19  &Ka97      &GL& 2.639& 0.68& 0.03& $<$0.5\,~~&     &0.4 &$-$1.2&1.7 &1.5& l \\ 
J0459$+$0229& 0457$+$024&V97O98    &Q & 2.384& 1.43& 0.06& $<$0.5\,~~&(0.1)&2.4 &$-$0.9&2.2 &1.1& c \\
J0503$+$0203& 0500$+$019&G83V97O98 &Q & 0.583& 2.05& 0.02& $<$0.5\,~~&(0.2)&2.8 &$-$0.6&1.4 &1.4& c \\
J0745$-$0044& 0743$-$006&V97O98    &Q & 0.994& 2.05& 0.04&      0.61&     &6.9 &$-$0.5&1.4 &1.2& c \\
J0900$-$2808& 0858$-$279&S85V97    &Q & 2.152& 3.18& 0.08&      0.52&     &5.2 &$-$1.0&1.5 &1.2& c \\ 
J1146$-$2447& 1143$-$245&V97O98    &Q & 1.950& 1.41& 0.04&      0.60&(2.1)&3.1 &$-$1.2&2.3 &1.3& c \\
J1357$-$1744& 1354$-$174&K97V97    &Q & 3.147& 1.16& 0.03&      4.92&     &1.9 &$-$0.6&1.0 &1.7& l \\
J1445$+$0958& 1442$+$101&V97O98    &Q & 3.535& 1.07& 0.03&      2.02&(2.0)&1.0 &$-$0.9&1.3 &1.3& c \\
J1522$-$2730& 1519$-$273&G83V97    &BL& 1.294& 1.74& 0.15&      2.45&(2.6)&5.8 &$-$0.5&1.0 &1.5& l \\
J1543$-$0757& 1540$-$077&V00       &G & 0.172& 1.00& 0.02& $<$0.5\,~~&     &0.7 &$-$0.7&0.8 &1.9& c \\
J1723$-$6500& 1718$-$649&T97       &G & 0.014& 4.64& 0.02& $<$0.5\,~~&     &2.7 &$-$0.6&1.1 &1.8& c \\
J1744$-$5144& 1740$-$517&K97       &G &  ... & 3.88& 0.02& $<$0.5\,~~&     &1.0 &$-$0.8&1.0 &1.7& c \\
J1939$-$6342& 1934$-$638&K97V97    &G & 0.183& 5.84& 0.01& $<$0.5\,~~&     &1.4 &$-$1.2&2.2 &1.0& c \\
J2003$-$3251& 2000$-$330&V97       &Q & 3.773& 0.92& 0.04& $<$0.5\,~~&     &5.7 &$-$1.7&2.6 &1.1& e \\
J2129$-$1538& 2126$-$158&K97V97O98 &Q & 3.270& 1.18& 0.02& $<$0.5\,~~&(0.4)&6.9 &$-$0.6&1.5 &1.3& l \\
J2136$+$0041& 2134$+$004&O98       &Q & 1.936& 9.47& 0.02&      0.60&(0.8)&5.9 &$-$0.6&2.2 &1.1& c \\
J2257$-$3627& 2254$-$367&T03a      &G & 0.006& 1.28& 0.04& $<$0.5\,~~&     &2.7 &$-$0.5&1.0 &1.8& c \\
J2336$-$5236& 2333$-$528&V97       &G & ...  & 1.63& 0.02& $<$0.5\,~~&     &1.1 &$-$0.6&0.7 &2.5& l \\
\noalign{\smallskip}
\hline
\end{tabular}
\begin{list}{}{}
\item[$^{\mathrm{a}}$] References:
G83 = \cite{gop83},
K97 = \cite{kin97},
Ka97 = \cite{kat97},
O98 = \cite{ode98},
S85 = \cite{spo85},
T97 = \cite{tin97},
T03a = \cite{tin03a},
V97 = \cite{dev97},
V00 = \cite{dev00}.
\end{list}
\end{table*}

\subsection{Notes on individual sources}

In this section we add supplementary notes or comments on individual sources:

\medskip
\noindent
J0240$-$2309 
has the second highest fractional polarization of this group.
VLBI imaging reveals a core-jet structure with the compact
core dominated at 2 and 5\,GHz by a steeper spectrum jet
extending to the north-east \citep{fey96,fom00}.

\medskip
\noindent
J0241$-$0815 (0238$-$084, NGC1052)  
is a low redshift galaxy, considered in detail by
\cite{kam01} and \cite{ver03}.
A spectrum compiled from an extensive list of sources is given by 
\cite{tor00}, revealing variability by a factor
of $\sim$2 over a wide range of frequencies. This variable
nature is confirmed by the fact that the
source has the second highest variability index in Table~\ref{tab1}.
The low frequency points lie above the nominal fit to the spectrum, and
are qualitatively consistent with the notion that this source has previously
been active, with the GPS spectrum reflecting a new recent stage of activity
\citep[e.g.,][]{tin03a}.

\medskip
\noindent
J0414+0534 (4C +05.19) is
a gravitational lens system.
The 178~MHz data point is from the 4C catalog 
\citep{gow67}.
A low variability index results from the ATCA monitoring, but
archival data shows significant variation,
ranging between 1.7 and 2.7\,Jy, at 1.4\,GHz.
Some of this variability may be due to micro-lensing, which has been suggested
as an explanation for the discrepant radio and optical flux ratios of 
the components \citep{wit95}.
The peak of our fitted spectrum is 0.4\,GHz, just below the nominal
GPS/CSS division.

\medskip
\noindent
J0459+0229: 
The 606\,MHz point is from \cite{kuh81}.
VLBI images reveal a core with a jet directed to the north
\citep[e.g.,][]{fey00,fom00}.

\medskip
\noindent
J0503+0203:
VLBI images of this CSO are given by
\cite{fey00}, \cite{fom00} 
and \cite{sta01}.

\medskip
\noindent
J0745$-$0044 
shows a pronounced turn up at low frequencies
with a flux density at 178~MHz
of 2.7\,($\pm$15\%)~Jy \citep{gow67}.
The spectral index at high frequencies is 
also flatter than most GPS sources.
A spectrum compiled from a much more extensive archival search
is given in \cite{tor00}.
A core dominated parsec-scale morphology is shown in the 
VLBA (pre-launch survey, VLBApls) image of \cite{fom00}.

\medskip
\noindent
J0900$-$2808:
The 80\,MHz flux density of 7\,Jy \citep{sle95} indicates a turn-up
at low frequencies.
The source is resolved out on 100\,M$\lambda$ baselines in the 
4.8\,GHz VLBA snapshot observation of \cite{fom00}.

\medskip
\noindent
J1146$-$2447:
The 2.1\% polarized flux reported by \cite{per82} is
significantly higher than the value of 0.6\% from the ATCA monitoring.
The VLBA observations of \cite{fom00} suggest
a core-jet morphology.

\medskip
\noindent
J1357$-$1744 
has the highest percentage linear polarization of this group.
The 43~GHz point suggests either a turn-up at high frequencies
(possibly associated with the emergence of the core at these
frequencies), or of variability.
At 4.8\,GHz, the compact parsec scale morphology was best fit by a 
bright central component with a fainter component $\sim$1\,mas
away on either side \citep{fom00}.

\medskip
\noindent
J1445+0958:
The 4.8\,GHz VLBA image of \cite{fom00}
shows a core and bent parsec-scale jet.

\medskip
\noindent
J1522$-$2730 
is classified as a BL Lac object, and, as noted in \S2, 
shows the highest variability index of this group.
The redshift for this object was recently reported by
\cite{hei04}.
Intra-Day Variability at radio wavelengths
has been observed in this source 
\citep{ked01,jau03}
and VLBI imaging confirms the core is very compact 
\citep[e.g.,][]{fom00}.
\cite{tor01} present a
composite spectrum derived from a large number of observations, 
and note that the source appears to have a 
relatively flat quiescent spectrum that is inverted during flares.
The high variability and apparent lack of a persistent GHz peaked
spectrum suggest this is less likely to be a bona fide GPS source.

\medskip
\noindent
J1543$-$0757:
The VLBA image reveals two extended components separated by 50\,mas, 
with the northern component $\sim$3 three times brighter 
\citep{fom00}.

\medskip
\noindent
J1744$-$5144: 
VLBI observations at 2.3\,GHz reveal a widely-spaced, extended
double with a separation of 52\,mas \citep{jau03}.

\medskip
\noindent
J1723$-$6500 
is a low-redshift galaxy,
considered in detail by  \cite{tin97,tin02,tin03a}.
A composite spectrum extending beyond 100\,GHz
is given by \cite{tor01}.
A detailed multi-frequency, multi-epoch study of the spectrum was undertaken
more recently by \cite{tdk03}.

\medskip
\noindent
J1939$-$6342 
is one of the earliest known GPS sources 
\citep{bol63,kel66}.
The spectrum has the narrowest width of this group (which is 
discussed further in \S7.2).
VLBI imaging reveals the
source has a compact double morphology with a separation of 42\,mas
\citep{tzi89,tzi98}.

\medskip
\noindent
J2003$-$3251 is the highest-redshift source in this group,
and has a
probable core-jet morphology on VLBI scales 
\citep{fom00}.

\medskip
\noindent
J2129$-$1538:
The \cite{kov99} 22\,GHz flux density of 1.374$\pm$0.021~Jy
does not lie on the spectrum defined by the other points.
This may imply some variability at higher frequencies,
which is also suggested by the spectrum compiled by 
\cite{tor01}.
The source has a core dominated parsec-scale morphology 
\citep{fom00}.

\medskip
\noindent
J2136$+$0041 (OX 057):
Another source recognised early on to be a GPS source
\citep{shi68}.
The low frequency points lie above (and were not used in deriving) the fitted spectrum, 
a trait
evident in the spectrum presented by \cite{kra68}.
This is also apparent in the spectrum compiled by
\cite{tor01}, which reveals that while the source
shows some variability, the GPS character of the spectrum
remains clearly present.
\cite{all02} present the centimetre wavelength variability over
a 30 year period, which reveals modest long-term variability.
\cite{lov00} describe VSOP space VLBI observations of the
source.  \cite{lis02} note the source had been
thought to be a symmetric double but that in fact it has a core-jet
morphology.

\medskip
\noindent
J2257$-$3627
was identified as a GPS source in this ATCA monitoring program
\citep{tin03a}.
As a low-redshift galaxy, high linear resolution studies over a range of
wavelengths have been possible, and, as discussed in 
\cite{tin03a},
there is strong evidence for merger activity in the galaxy which is quite
likely to be related to the GPS spectrum.

\medskip
\noindent
J2336$-$5236 has
the broadest spectrum and smallest spectral curvature of the
sources in this table.
The source was originally identified as a quasar, however this was called
into doubt \citep{jau89}, and more recently it has been
identified as a galaxy \citep{dev95}.

\section{Newly identified GPS sources and candidates}

\begin{figure*}
 \centering
 \includegraphics{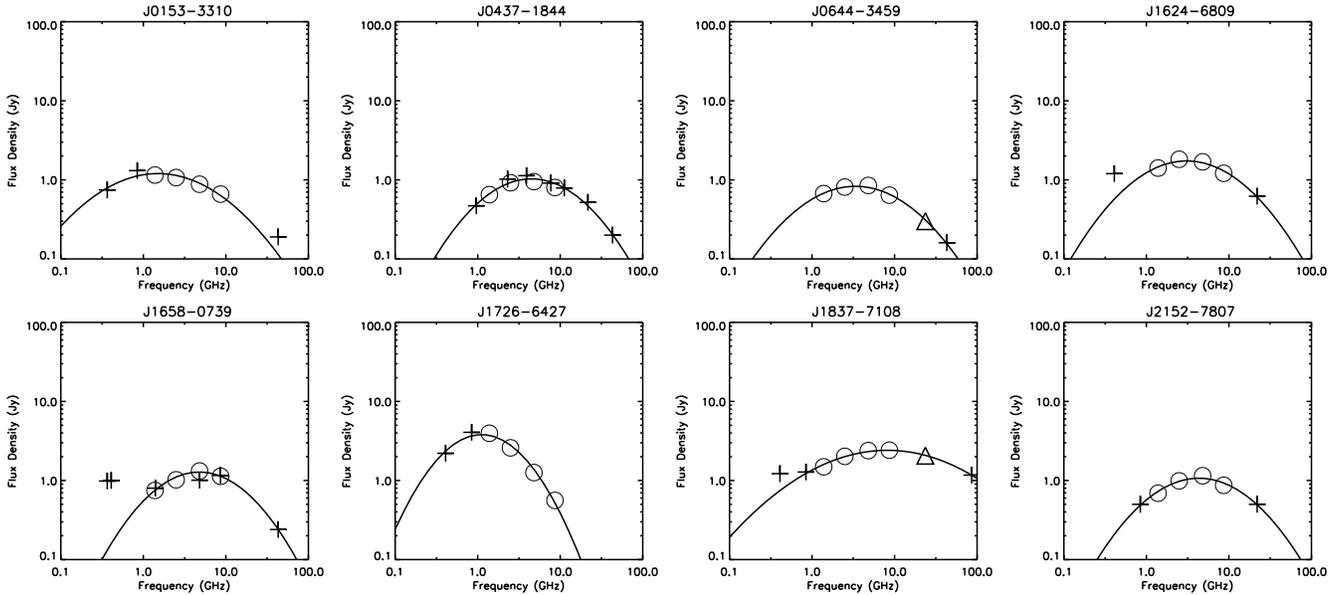}
 \caption{Spectra of newly identified GPS sources.
    Symbols are the same as for Figure~1.
    The fit to the spectra is described in \S2 and
    results given in Table~\ref{tab2}.
 }
 \label{Fig2}%
 \end{figure*}

In Table~\ref{tab2} the properties of eight new candidate GPS sources are 
presented, with the spectra plotted in Figure~\ref{Fig2}.  
The primary requirements for selection were
a spectral peak above 500~MHz, a spectral index at high
frequencies of $\alpha_{hi} < -0.5$, and a spectral curvature of at least
0.6. 
As we found that the 
variability indices and fractional polarizations were lower 
for previously reported GPS sources, we were also guided in our
selection by these parameters.

%____________________________________ Two column figure (place early!)

%__________________________________________________ One column table
\begin{table*}
\caption[]{Newly identified GPS sources and candidates.
Type EF denotes empty field. Other types and parameters are 
as for Table~\ref{tab1}.
}
\label{tab2}
\begin{center}
\begin{tabular}{l l c c c c c rl c c c c}
\hline
\noalign{\smallskip}
Source & B1950 & Type & $z$ & $S_{4.8}$ & $m_{4.8}$ & \multicolumn{2}{c}{$p_{4.8}$} & $\nu_{pk}$& $\alpha_{hi}$ & $\Delta\alpha $ &width & flag \\
 &  &  & & (Jy) &  & \multicolumn{2}{c}{(\%)} & (GHz) &  &  & &  \\
\noalign{\smallskip}
\hline
\noalign{\smallskip}
J0153$-$3310& 0150$-$334&Q &0.610& 0.88& 0.04&      1.35&(1.1)&1.5&$-$0.7&1.2&1.6& e \\
J0437$-$1844& 0434$-$188&Q &2.702& 0.95& 0.05&      0.80&(0.8)&4.5&$-$1.2&2.1&1.3& l \\
J0644$-$3459& 0642$-$349&Q &2.165& 0.85& 0.09&      2.03&(1.1)&3.3&$-$1.0&1.3&1.4& c \\
J1624$-$6809& 1619$-$680&Q &1.360& 1.69& 0.04&$<$0.5\,~~&     &3.1&$-$0.7&1.1&1.4& c \\
J1658$-$0739& 1656$-$075&EF&...  & 1.32& 0.03&$<$0.5\,~~&     &4.8&$-$1.0&1.5&1.2& e \\
J1726$-$6427& 1722$-$644&EF& ... & 1.26& 0.02&$<$0.5\,~~&     &1.1&$-$1.4&2.2&1.0& c \\
J1837$-$7108& 1831$-$711&Q &1.356& 2.39& 0.06&      1.77&     &8.2&$-$0.4&0.8&2.0& c \\
J2152$-$7807& 2146$-$783&Q & ... & 1.15& 0.03&$<$0.5\,~~&     &4.3&$-$0.6&1.2&1.3& l \\
\noalign{\smallskip}
\hline
\end{tabular}
\end{center}
\end{table*}

\subsection{Notes on individual sources}

\medskip
\noindent
J0153$-$3310 
has a low variability index in the ATCA monitoring, but archival
data suggest it is moderately variable on longer timescales,
with 5\,GHz flux densities ranging from 0.860~Jy 
in the PKS Catalogue (PKSCAT) \citep{wri90} to 1.378~Jy in the
PMN survey \citep{wri96}.
The VLBA snapshot image of \cite{fom00} shows a core-dominated
structure that was best modelled by two components of similar flux density
separated by 1\,mas.

\medskip
\noindent
J0437$-$1844: 
VLBI imaging has been undertaken by \cite{fey00} and 
\cite{fom00}: 
the latter observation yielded a core-dominated
structure modelled by two components of comparable flux density
separated by 1\,mas.

\medskip
\noindent
J0644$-$3459: 
The 4.8\,GHz VLBA image of \cite{fom00} is comprised of a core and a
more extended (jet) component 3\,mas to the west.

\medskip
\noindent
J1624$-$6809: 
Including the 408\,MHz point in the fit to the spectrum
would increase the FWHM from 1.4 to 1.7.
The 843\,MHz SUMSS flux density, when available, will be helpful
in determining whether the spectrum is indeed intrinsically broader,
or whether it is narrower but with a turn-up at lower frequencies.
\cite{bea97} give 5$\sigma$ upper limits at 89\,GHz and
147\,GHz of 0.08\,Jy and 0.05\,Jy, respectively.

\medskip
\noindent
J1658$-$0739:
The ATCA and VLA data points clearly yield a peaked spectrum,
and the few archival flux density measurements confirm the
source is not very variable.
The 365 and 408\,MHz data points indicate a spectral turn-up at lower 
frequencies, suggestive of extended emission,
and consistent with the `e' flag for this source.
The low variability and polarization
properties from the ATCA monitoring match those of canonical
GPS sources very closely.
The 4.8\,GHz VLBA image of \cite{fom00} is dominated by the core,
with a weak secondary 7\,mas away.

\medskip
\noindent
J1726$-$6427 
has the narrowest spectrum of the sources in Table~\ref{tab2}, similar
to that of PKS 1934$-$638 (see \S7.2), but otherwise currently
little is known about this source.

\medskip
\noindent
J1837$-$7108:
The spectrum is relatively broad,
with a resulting spectral index between 22 and 89\,GHz of $-$0.4,
however the 146\,GHz flux density of 0.73\,Jy 
\citep{bea97}
indicates the spectrum steepens, with a spectral index of $-$0.95
between 89 and 146\,GHz.
The PKSCAT flux density of 1.15\,Jy at 5.0\,GHz 
\citep{wri90} is roughly half that
of the corresponding values from the PMN survey, 2.29\,Jy 
\citep{wri94}, 
and the ATCA monitoring, 2.39\,Jy.  Similarly, the PKSCAT
flux density at 2.7\,GHz of 1.32\,Jy is $\sim$65\% of the 2.02\,Jy in 
the ATCA monitoring, suggesting the source has brightened significantly
over the intervening period.

\medskip
\noindent
J2152$-$7807:
Identified as a quasar by \cite{jau89}.
Comparison with catalogued flux densities implies the source is
moderately variable on long timescales.  
The PMN flux density of 1.128\,Jy 
\citep{wri94} is in
good agreement with the ATCA monitoring value of 1.15\,Jy, with both
being significantly in excess of the PKSCAT flux density of 0.77\,Jy
\citep{wri90}.
\cite{bea97} give 5$\sigma$ upper limits at 3\,mm (89\,GHz)
and 2\,mm (147\,GHz) from SEST observations of 0.06 and 0.11\,Jy,
respectively.

 \begin{figure*}
 \centering
 \includegraphics{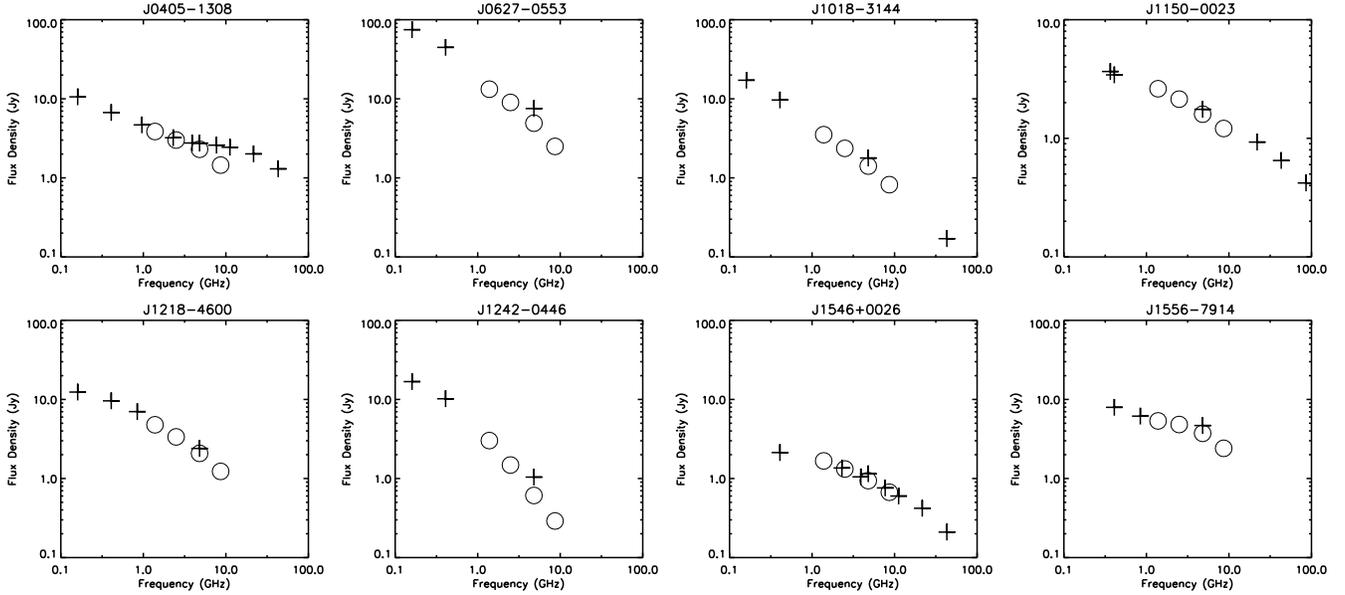}
 \caption{Spectra of previously identified CSS sources.
    Symbols are the same as for Figure~\ref{Fig1}.
  }
 \label{Fig3}%
 \end{figure*}

\section{Previously identified CSS sources and candidates}

The properties of previously reported CSS sources are listed
in Table~\ref{tab3} and their spectra are plotted in Figure~\ref{Fig3}.
The spectral indices, $\alpha$,  given in Tables~\ref{tab3} 
and~\ref{tab4} are determined 
between 4.8 and 8.4\,GHz from the ATCA monitoring \citep{tin03b}.

%__________________________________________________ One column table
\begin{table*}
\caption[]{Properties of previously identified CSS sources and candidates.
Parameters are the same as for Tables~\ref{tab1} and~\ref{tab2}.
}
\label{tab3}
\begin{center}
\begin{tabular}{l l l c c c c rl c c}
\hline
\noalign{\smallskip}
Source & B1950 & Refs\,$^{\mathrm{a}}$ & Type & $z$ & $S_{4.8}$ & $m_{4.8}$ & \multicolumn{2}{c}{$p_{4.8}$} & $\alpha$ & flag \\
 &  & &  &  & (Jy) &  & \multicolumn{2}{c}{(\%)} &  &  \\
\noalign{\smallskip}
\hline
\noalign{\smallskip}
J0405$-$1308& 0403$-$132& K97 &Q & 0.571& 2.30& 0.05&   2.40   &     & $-$0.79& e \\
J0627$-$0553& 0624$-$058& K97 &EF&  ... & 4.92& 0.03&   6.53   &     & $-$1.15& e \\
J1018$-$3144& 1015$-$314& K97 &Q & 1.246& 1.41& 0.02&$<$0.5\,~~&(0.1)& $-$0.92& c \\
J1150$-$0023& 1148$-$001& K97 &Q & 1.983& 1.60& 0.03&   2.88   &     & $-$0.53& l \\ 
J1218$-$4600& 1215$-$457& K97 &Q & 0.529& 2.08& 0.03&   2.23   &     & $-$0.88& e \\
J1242$-$0446& 1239$-$044& M92 &G & 0.480& 0.61&  ...&   ....   &     & $-$1.20& e \\
J1546$+$0026& 1543$+$005& G83 &G & 0.550& 0.94& 0.05&$<$0.5\,~~&     & $-$0.59& l \\
J1556$-$7914& 1549$-$790& K97 &G & 0.150& 3.73& 0.02&$<$0.5\,~~&     & $-$0.74& c \\
\noalign{\smallskip}
\hline
\end{tabular}
\end{center}
\begin{list}{}{}
\item[$^{\mathrm{a}}$] References:
G83 = \cite{gop83}, 
K97 = \cite{kin97},
M92 = \cite{man92}.
\end{list}
\end{table*}

It is notable that, in contrast to the GPS sources and candidates in
Tables~\ref{tab1} and~\ref{tab2}, a number of the CSS sources have `e' flags, indicating
evidence of extended structure on 6\,km ATCA baselines. This is
not unexpected, as the angular resolution at 4.8\,GHz on this
baseline is 2$''$, which corresponds to a linear resolution of about
10\,kpc for a source at $z\sim0.5$.  These previously reported
CSS sources have, like the previously reported GPS sources,
low variability indices.

\subsection{Notes on individual sources}

\medskip
\noindent
J0405$-$1308:
The 80\,MHz flux density of \cite{sle95} is 11\,Jy, indicating
that any turnover, if one exists, lies at lower frequencies.
The higher flux density points from \cite{kov99}, together
with higher frequency points in the VLA catalog, suggest a flattening,
or variability, of the spectrum above $\sim$6\,GHz.
The snapshot VLBA image of \cite{fom00} contains
a very compact core and fainter jet extending to the south-east.

\medskip
\noindent
J0627$-$0553 (3C161):
The 80\,MHz flux density of \cite{sle95} is 111\,Jy.
\cite{per82} notes significant extended structure on VLA baselines,
and the source was not detected in the VLBA observations of 
\cite{fom00}.

\medskip
\noindent
J1018$-$3144:
The source was highly
resolved in the VLBA snapshot observation of \cite{fom00}, 
with a correlated flux density of 0.5~Jy detected only on the
shortest baselines.

\medskip
\noindent
J1150$-$0023 (PMN J1150$-$0024):
Referred to as a GPS source by \cite{jau03}, however 
it would appear any peak lies below 350\,MHz and so we classify it here
as a CSS source.
\cite{sle95} gives 80 and 160\,MHz flux densities of 5\,Jy and 3.0\,Jy
respectively.
The snapshot VLBI observation of 
\cite{fom00} yielded a core-jet morphology,
with the jet extending over $\sim$30\,mas form the core.

\medskip
\noindent
J1218$-$4600:
The 80\,MHz flux density of \cite{sle95} is 16\,Jy.

\medskip
\noindent
J1242$-$0446 (3C275):
Dropped from the ATCA monitoring after the first few epochs and so no 
variability 
index or polarization information is given in \cite{tin03b}.
VLA observations at 8.4 and 15 GHz did not reveal any nuclear component
\citep{man97}
and, as suggested by that result, the source was
not detected in the 4.8\,GHz VLBA observations of 
\cite{fom00}.

\medskip
\noindent
J1546+0026:
Emission is seen on both sides of the core in the images of
\cite{sta99} 
and \cite{fom00}, and also 
\cite{pec00}, who classify the source as a CSO.

\medskip
\noindent
J1556$-$7914:
The observations of \cite{lov97} confirm the compact nature
of the source on ATCA baselines, with 99\% of the source flux density
contained within an unresolved core component at 4.8\,GHz.

 \begin{figure*}
 \centering
 \includegraphics{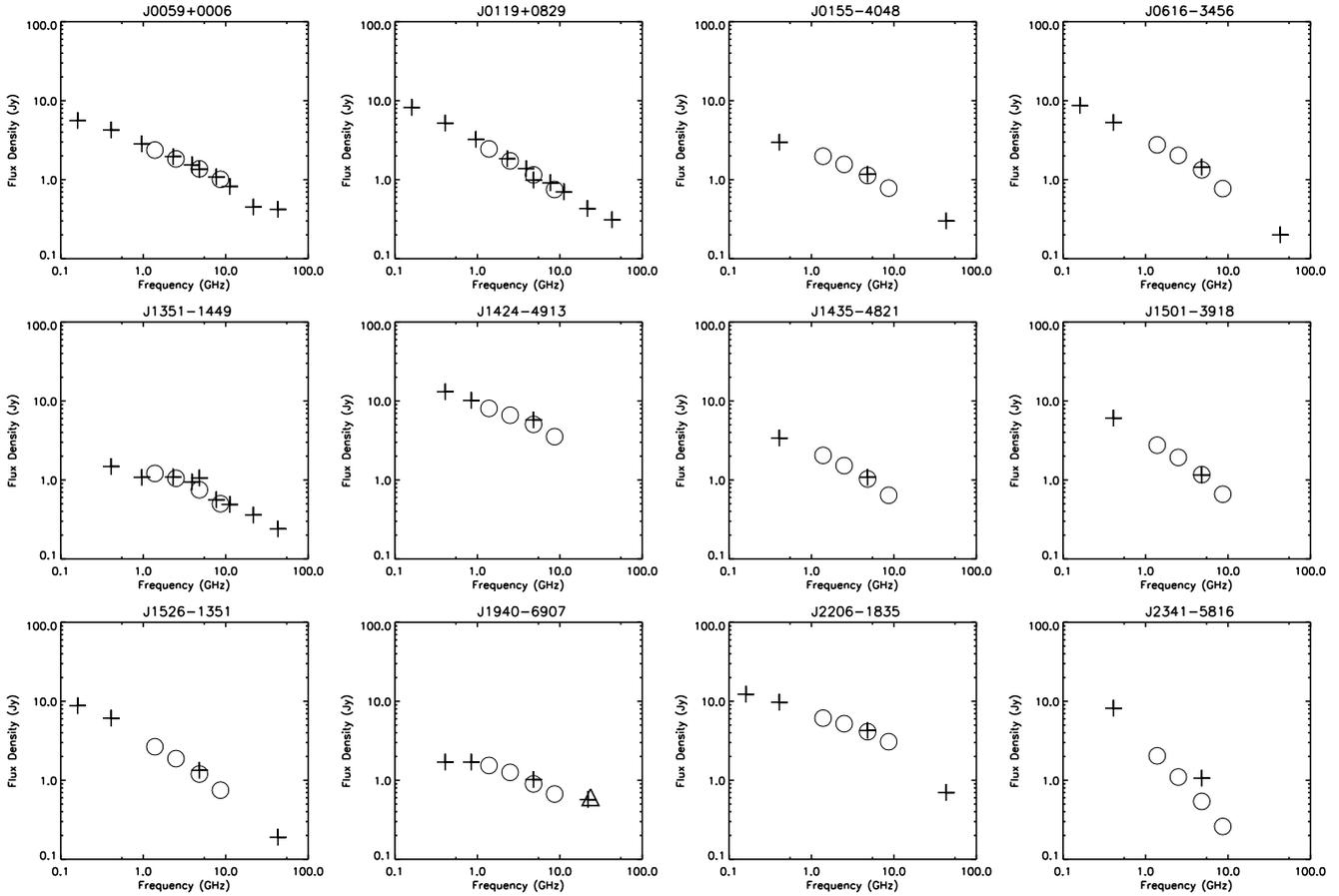}
 \caption{Spectra of newly identified CSS sources.
    Symbols are the same as for Figure~\ref{Fig1}.
 } 
 \label{Fig4}%
 \end{figure*}

\section{Newly identified CSS sources and candidates}

The twelve CSS sources and candidates found in the ATCA monitoring
observations are listed in Table~\ref{tab4} and their spectra are plotted
in Figure~\ref{Fig4}. 
For selection, spectra were
required to have 
a spectral index ($S \propto \nu^{+\alpha}$) 
between 4.8 and 8.4\,GHz  of less than $\sim -0.5$
and a peak, if any, below $\sim$500~MHz,
and again we were also guided by the variability index
and fractional polarization.
Two sources have spectra indices marginally below the nominal
cut-off of, e.g., \cite{fan90}, 
however we include them as candidates worthy of
further study, as the nominal cut-off is purely a convention and
does not have any particular physical meaning.

The variability indices are consistent
with the low values observed in the previously reported 
CSS sources. 
Higher resolution observations are required to determine whether 
a number of these candidates satisfy the compactness 
requirement for CSS sources.

%__________________________________________________ One column table
\begin{table*}
\caption[]{Newly identified CSS sources and candidates.
Parameters are the same as for previous Tables.
}
\label{tab4}
\begin{center}
\begin{tabular}{l l c c c c rl c c}
\hline
\noalign{\smallskip}
Source  & B1950   &  Type & $z$ & $S_{4.8}$ & $m_{4.8}$ & \multicolumn{2}{c}{$p_{4.8}$} & $\alpha$ & flag \\
        &         &       &     & (Jy)  &       & \multicolumn{2}{c}{(\%)}  &          &       \\
\noalign{\smallskip}
\hline
\noalign{\smallskip}
J0059$+$0006 & 0056$-$001 & Q & 0.719 & 1.37&  0.02&    7.25  &(5.8)&  $-$0.52& c \\
J0119$+$0829 & 0116$+$082 & G & 0.594 & 1.15&  0.02&    1.60  &     &  $-$0.72& c \\
J0155$-$4048 & 0153$-$410 & G & 0.226 & 1.13&  0.01&    0.82  &     &  $-$0.63& c \\
J0616$-$3456 & 0614$-$349 & G & 0.329 & 1.33&  0.02&    0.61  &     &  $-$0.93& c \\
J1351$-$1449 & 1349$-$145 & EF&  ...  & 0.75&  0.05& $<$0.5\,~~&    &  $-$0.67& c \\
J1424$-$4913 & 1421$-$490 & EF&  ...  & 5.06&  0.03&    3.22  &     &  $-$0.61& e \\
J1435$-$4821 & 1431$-$481 & EF&  ...  & 1.03&  0.02&    1.01  &     &  $-$0.81& c \\
J1501$-$3918 & 1458$-$391 & EF&  ...  & 1.17&  0.02& $<$0.5\,~~&    &  $-$0.97& c \\
J1526$-$1351 & 1524$-$136 & Q & 1.687 & 1.20&  0.02&    2.56  &(2.8)&  $-$0.80& e \\
J1940$-$6908 & 1935$-$692 & Q & 3.154 & 0.90&  0.09&    2.01  &     &  $-$0.49& l \\ 
J2206$-$1835 & 2203$-$188 & Q & 0.620 & 4.12&  0.02&    2.67  &(3.0)&  $-$0.49& e \\
J2341$-$5816 & 2338$-$585 & EF&  ...  & 0.54&  ... &    ...   &     &  $-$1.18& e \\
\noalign{\smallskip}
\hline
\end{tabular}
\end{center}
\end{table*}

\subsection{Notes on individual sources}

\medskip
\noindent
J0059$+$0006 has a
relatively high fractional polarization: only four sources have 
$p_{4.8} > $7\%.
The 4.8\,GHz VLBA observation of \cite{fom00} revealed
a compact core with jet emission extending beyond
20\,mas from the core.

\medskip
\noindent
J0119$+$0829:
The snapshot VLBA images shows a curved jet extending over 
to the south with a bright component 30\,mas north of the core
\citep{fom00}.

\medskip
\noindent
J0155$-$4048:
\cite{lov97} confirms the compact nature of the source on
6\,km baselines, finding that 100\% of the source flux density is
contained within an unresolved core component at 4.8\,GHz.  The source
is mostly resolved out on the shortest VLBA baselines 
\citep{fom00}. Emission appears on both sides of the core,
suggestive of a CSO morphology, with the source extending over
$\sim$20\,mas.

\medskip
\noindent
J0616$-$3456 
was resolved
with the VLA to a 2$''$ source at 5\,GHz \citep{ulv81} and
was not detected in the VLBA observations of \cite{fom00}.

\medskip
\noindent
J1351$-$1449: 
A borderline GPS/CSS source, with the data hinting at a spectral
peak around 500\,MHz.
A core dominated source in the VLBA observations 
of \cite{fom00},
with diffuse emission centred 20\,mas north of the core.

\medskip
\noindent
J1424$-$4913:  
\cite{lov97}
found that 95\% of the source flux density is contained within
an unresolved core component on ATCA 6\,km baselines at 4.8\,GHz.

\medskip
\noindent
J1501$-$3918 has a 
strong core with possible short jet and a second component $\sim$130\,mas
from the core in the VLBA image of \cite{fom00}.

\medskip
\noindent
J1526$-$1351
(PMN J1526$-$1350): 
The source was mostly resolved out on 
25\,M$\lambda$ baselines at 4.8\,GHz \citep{fom00}.

\medskip
\noindent
J1940$-$6907 (PMN J1940$-$6908):
The highest frequency points suggest a flattening of the spectrum
at high frequencies. This would result in the source being
rejected as a CSS candidate based on the strictest selection criteria,
but may indicate the emerging visibility of
a flat or inverted spectrum core.
VLBI observations are required to determine the parsec-scale
morphology.

\medskip
\noindent
J2206$-$1835:
The source was
resolved out on 10\,M$\lambda$ baselines at 4.8\,GHz 
\citep{fom00}.

\medskip
\noindent
J2341$-$5816: 
Dropped from the ATCA monitoring after the first few epochs 
so no variability index or fractional
polarization is given by \cite{tin03b}.

 \begin{figure*}
 \centering
 \includegraphics{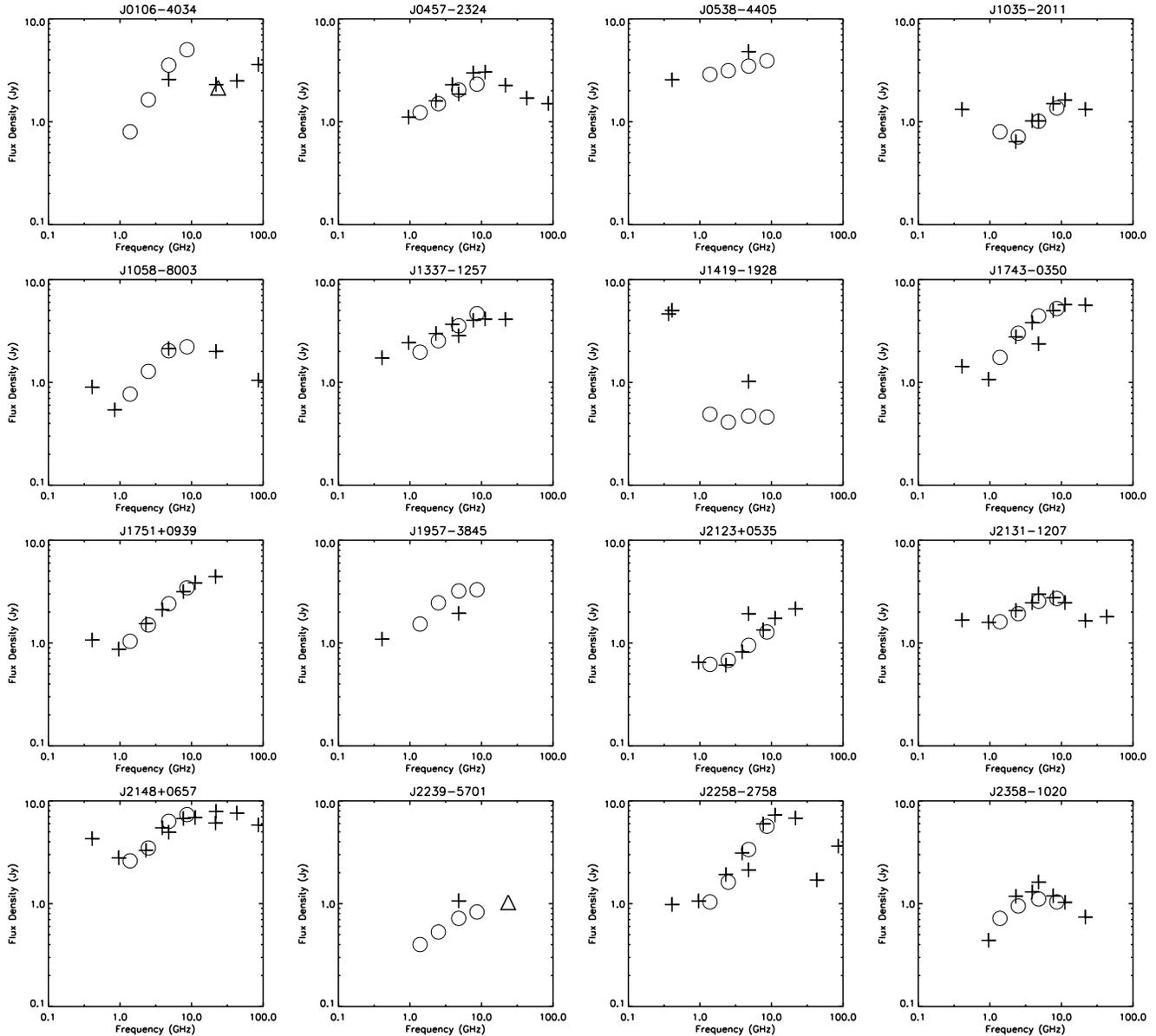}
 \caption{Spectra of other sources of interest.
    Symbols are the same as for Figure~\ref{Fig1}.
 }
 \label{Fig5}%
 \end{figure*}

\section{Other sources of interest}

A number of other sources have been suggested as GPS or CSS candidates in
the literature. However, as discussed in this section, they do not
meet one of more of the accepted criteria for GPS or CSS sources. 
The ATCA monitoring program included 7 of the 12 
``new sources with GPS-type spectra'' 
and 7 of the 8 ``other sources with inverted spectra''
given by \cite{tor01}. As noted by 
\cite{tor01}, many of these
sources show evidence for significant variability, and this
is reflected by the fact that 11 of the 14 sources had variability
indices at 4.8\,GHz (derived from the ATCA monitoring program)
above the median value. For many of these sources the
spectra based on the average fluxes over the epochs of the ATCA monitoring 
did not meet our criteria for GPS or CSS sources.
\cite{tor01} point out that many of the sources they 
consider have higher spectral peaks than most catalogued GPS sources, and so
some of these may be related to the class of High Frequency Peakers (HFPs),
which are  somewhat more variable than GPS/CSS sources
\citep{dal00,dal02,dal03}.
Table~\ref{tab5} and Figure~\ref{Fig5} 
contain a number of the sources from \cite{tor01}, 
and in addition a number of sources from our ATCA monitoring
with inverted spectra up to 8.4\,GHz --- candidate HFPs.
We also include J1419$-$1928,
designated as a candidate CSO by \cite{tay03}.

%__________________________________________________ One column table
\begin{table*}
\caption[]{Other sources of interest. 
Sources for which no reference is given are those from the ATCA
program with spectra that are inverted, but which do not meet our
criteria for inclusion as GPS candidates.
The spectral index, $\alpha$, is that determined between 4.8 and 
8.4\,GHz from the ATCA monitoring \citep{tin03b}.
Parameters are the same as for previous Tables.
}
\label{tab5}
\begin{center}
\begin{tabular}{l l l c c c c rl r c}
\hline
\noalign{\smallskip}
Source & B1950 & Refs\,$^{\mathrm{a}}$  & Type & $z$ & $S_{4.8}$ & $m_{4.8}$ & \multicolumn{2}{c}{$p_{4.8}$} & $\alpha$~ & flag \\
 &  &  &  &  & (Jy) &  & \multicolumn{2}{c}{(\%)} &  &  \\
\noalign{\smallskip}
\hline
\noalign{\smallskip}
 J0106$-$4034& 0104$-$408&    & Q & 0.584& 3.56& 0.06&   1.34   &(1.4)&   0.57& c \\ 
 J0457$-$2324& 0454$-$234& T01& Q & 1.003& 2.04& 0.15&   1.44   &(3.5)&   0.23& l \\
 J0538$-$4405& 0537$-$441& T01& BL& 0.894& 3.48& 0.30&   1.28   &(1.0)&   0.15& l \\
 J1035$-$2011& 1032$-$199&    & Q & 2.198& 1.01& 0.11&   3.03   &(4.1)&   0.52& l \\ 
 J1058$-$8003& 1057$-$797& T01& Q & ...  & 2.03& 0.20&   3.14   &     &   0.16& c \\ 
 J1337$-$1257& 1334$-$127& T01& Q & 0.539& 3.55& 0.24&   3.34   &     &   0.41& l \\
 J1419$-$1928& 1417$-$192& T03& G & 0.120& 0.47& 0.07&   2.44   &     &$-$0.03& e \\
 J1743$-$0350& 1741$-$038&    & Q & 1.054& 4.43& 0.11&   0.96   &(1.2)&   0.29& c \\ 
 J1751$+$0939& 1749$+$096&    & BL& 0.322& 2.41& 0.28&   1.41   &(6.1)&   0.61& c \\ 
 J1957$-$3845& 1954$-$388& T01& Q & 0.626& 3.21& 0.26&   3.09   &(1.0)&   0.05& c \\
 J2123$+$0535& 2121$+$053&    & Q & 1.878& 0.95& 0.32&   1.17   &(2.5)&   0.54& c \\ 
 J2131$-$1207& 2128$-$123& T01& Q & 0.499& 2.55& 0.04&   1.20   &(0.6)&   0.09& c \\
 J2148$+$0657& 2145$+$067& T01& Q & 0.990& 6.33& 0.05&$<$0.5\,~~&(1.0)&0.25&c \\
 J2239$-$5701& 2236$-$572&    & EF& ...  & 0.72& 0.05&   4.39   &     &   0.24& c \\ 
 J2258$-$2758& 2255$-$282& T01& Q & 0.926& 3.37& 0.10&   0.92   &     &   0.89& l \\
 J2358$-$1020& 2355$-$106&    & Q & 1.626& 1.11& 0.34&   2.84   &     &$-$0.18& c \\ 
\noalign{\smallskip}
\hline
\end{tabular}
\end{center}
\begin{list}{}{}
\item[$^{\mathrm{a}}$] References:
T01 = \cite{tor01},
T03 = \cite{tay03}.
\end{list}
\end{table*}

It is notable that the sources here generally have higher
variability indices and fractional polarizations than those
in the preceding tables.
In addition, larger differences are evident in the 
fractional polarizations measured by \cite{per82} and those 
measured in the ATCA monitoring.
The tendency for increased
variability in sources with more inverted spectra 
\citep[e.g.,][]{kes77,tin03b} is seen in the variability indices:
in the sample of 185 sources only 8 (4.3\%) have variability
indices at 4.8\,GHz of 0.30 or higher, with three
of these appearing in Table~\ref{tab5}.

\subsection{Notes on individual sources}

\medskip
\noindent
J0106$-$4034
has the fourth highest 4.8 to 8.4\,GHz spectral index in the ATCA sample.
PKSCAT flux densities of 0.57\,Jy at 2.7\,GHz and 0.85\,Jy at 5.0\,GHz 
differ significantly from corresponding values of 1.64 and 3.56\,Jy
from the ATCA monitoring. 
The PMN 4.85\,GHz flux density
of 2584$\pm$99\,mJy \citep{wri94}
and high frequency flux densities evident in Fig.~\ref{Fig5}
also suggest significant variability, although the variability index
is quite moderate. \cite{ked01} 
report evidence for intra-day variability at 2.4\,GHz.

\medskip
\noindent
J0457$-$2324: 
Noted as a new source with a GPS-type spectrum by 
\cite{tor01},
with turnover at 3\,GHz.
Possible core-jet morphology
in the VLBA observations of \cite{fom00}.

\medskip
\noindent
J0538$-$4405: 
Noted as a new source with a GPS-type spectrum by 
\cite{tor01}, with spectral peak at 5~GHz.

\medskip
\noindent
J1035$-$2011:
A turn-up at low frequencies is apparent. The 4.8\,GHz image of 
\cite{fom00} shows a core-jet morphology.

\medskip
\noindent
J1058$-$8003: 
Identified as a quasar by \cite{jau89}.
Noted as a new source with a GPS-type spectrum by 
\cite{tor01}, with a
peak at 10\,GHz, and the spectrum in Fig.~\ref{Fig5} is consistent with
this. \cite{bea97} measured a flux density of
0.41\,Jy at 147\,GHz, indicating the spectrum steepens further at high
frequencies. The turn-up at low frequencies notwithstanding, the
spectrum of the relatively sparsely observed source is consistent with
that of GPS sources. However, as the variability indices at all four
frequencies of the ATCA monitoring were relatively high
we do not include it as a candidate GPS source, 
although it is worthy of further study.

\medskip
\noindent
J1337$-$1257: 
Noted as a new source with a GPS-type spectrum by 
\cite{tor01}, with turnover at 20\,GHz.
The ATCA data have a spectral index of $\sim$0.4, though the data of 
\cite{kov99} suggest the spectrum is not as inverted as this.
Very compact in the VLBA observations of 
\cite{fom00}.

\medskip
\noindent
J1419$-$1928:
This galaxy has been noted as a candidate CSO by \cite{tay03}
based on their 5 and 15\,GHz imaging. As noted by 
\cite{fas01},
while there is some overlap between CSOs (defined by their morphology)
and GPS/CSS sources (defined largely by their spectra) there is
no clear correspondence between the two classes.
Certainly the spectrum based on the ATCA data in Fig.~\ref{Fig5} would not invite 
classification as either a GPS or CSS source. The higher PMN flux
density of 1.021~Jy at 4.85~GHz \citep{gri94}, 
together with the `e' structure flag, suggest the source
is partially resolved out on the 6\,km ATCA baseline. Excluding
the ATCA points yields a CSS-like spectrum with a spectral index
of $-$0.6.

\medskip
\noindent
J1743$-$0350: 
The variability index at 4.8\,GHz is moderate, however the
ATCA 4.8\,GHz flux density of 4.43\,Jy is significantly
greater than the 2.37\,Jy of the PMN \citep{gri95}.
VSOP space VLBI observations are described by \cite{waj00}.

\medskip
\noindent
J1751$+$0939:
This source has the third most inverted spectrum of the
ATCA sample of 202 sources, with a 4.8 to 8.4\,GHz spectral
index of 0.61. 
The fractional polarization of 6.1\% measured by \cite{per82} 
is much higher than the ATCA value of 1.41\%. 
There is a pronounced turn-up at low frequencies.

\medskip
\noindent
J1957$-$3845: 
Noted as a new source with a GPS-type spectrum by 
\cite{tor01}, with turnover at $\sim$10~GHz.
The compilation of data in Fig.~\ref{Fig5} would suggest that the 
spectral peak lies closer to 8\,GHz. 
A compact core-jet morphology was revealed by
the VLBA observations of \cite{fom00}.

\medskip
\noindent
J2123$+$0535:
The spectral index between 4.8 and 8.43\,GHz, 0.54, is the
fifth highest of the ATCA sample. 
In addition, the PMN flux density of 1.93\,Jy \citep{gri95} 
is almost double the ATCA value.
The VLBA observations of \cite{fom00} revealed a 
bright core and a compact secondary component 1.2\,mas away.

\medskip
\noindent
J2131$-$1207: 
Noted as a new source with a GPS-type spectrum by 
\cite{tor01}, with turnover at 5~GHz.
The 80\,MHz flux density of 6\,Jy \citep{sle95} indicates a large
turn-up at low frequencies.
Compact core-jet morphology in the VLBA observations of \cite{fom00}.

\medskip
\noindent
J2148$+$0657: 
Noted as a source with an inverted spectrum by 
\cite{tor01}.
The compilation of data in Fig.~\ref{Fig5} reveals a turn-up at low frequencies
and a peak around 30\,GHz. The variability index is quite low,
however, as noted by 
\cite{tor01}, over longer timescales
the source is strongly variable.
The fractional polarization upper limit is consistent 
with those of GPS/CSS sources. 
A compact core-jet morphology was seen in the VLBA observations of 
\cite{fom00}.

\medskip
\noindent
J2239$-$5701:
A relatively poorly studied source
with a low variability index in the ATCA monitoring.
The ATCA 4.8\,GHz flux density lies between the 
PKSCAT value of 0.49\,Jy \citep{wri90} and
the PMN value of 1.063$\pm$0.056 
\citep{wri94}, indicating
moderate longer term variability.
Our 22\,GHz measurement in April 2003 suggests the spectral
peak lies (for that epoch, at least) above 20\,GHz.

\medskip
\noindent
J2258$-$2758: 
Noted as a new source with a GPS-type spectrum by 
\cite{tor01}, with a peak at 20~GHz.
The 4.8 to 8.4\,GHz spectral index was the highest of the ATCA sources.
The variability index is only just above the median, however
the spectra compiled by 
\cite{tor01} reveals variations of almost
an order of magnitude in flux densities at mm wavelengths.
A compact core-jet morphology was seen in 
the VLBA observations of \cite{fom00}.

\medskip
\noindent
J2358$-$1020 has the most GPS-like spectrum of the sources
in Figure~\ref{Fig5}. The spectral index steepens from $-$0.18 
between 4.8 and 8.6\,GHz (ATCA data) to $-$0.50 between 11.2 and 21.65\,GHz
\citep{kov99}, consistent with the requirements for a GPS source. However,
as the source has one of the largest variability indices
at 4.8 and 8.6\,GHz in the ATCA monitoring, we do not include it
as a candidate. Variability is also apparent on longer timescales, as
the ATCA 4.8\,GHz flux
density is intermediate between the 
0.63\,Jy PKSCAT value \citep{wri90} and the
1.618$\pm$0.085~Jy of PMN \citep{gri94}.
Furthermore, inspection of Figure~2 of \cite{tin03b} indicates 
that the spectrum 
changes significantly over time, with spectral indices of
$\sim$0.0 all the way from 1.4 to 8.6\,GHz at some epochs.
The 4.8\,GHz VLBA observations of 
\cite{fom00} yielded
a very compact, core-dominated source.

\section{Discussion}

\subsection{Incidence of GPS and CSS sources}

There are 28 GPS sources and candidates in Tables~\ref{tab1} and~\ref{tab2},
constituting 14\% of the sample of 202 sources, in 
general agreement
with the $\sim$10\% incidence rate given by \cite{ode98}.
Tables~\ref{tab3} and \ref{tab4} contain 20 CSS sources, 10\% of the sample.
This is, as expected, significantly lower than the $\sim$30\% given by
\cite{ode98}, as there was a bias toward flat-spectrum sources
in the compilation of the 202 sources.

 \begin{figure*}
 \centering
 \includegraphics[angle=-90,width=8cm]{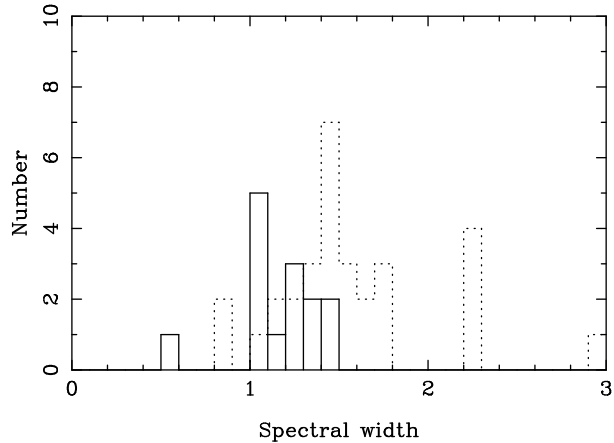}
 \caption{Spectral width (FWHM in decades of frequency) 
    for the CD/CSO sources (solid
    line histogram) and other sources (dotted line histogram) from the 
    sample of 44 sources of \cite{sne00}.
 }
 \label{Fig6}%
 \end{figure*}

\subsection{Spectral width}

In the sample of 15 sources of \cite{ode91},
0108+388 had the narrowest spectrum
with a FWHM of $\sim$0.95.
O'Dea et al.\ note that a width of 0.77 decades of frequency
is close to the minimum for an ideal source --- a homogeneous
self-absorbed synchrotron source with a power law electron
energy spectrum. The presence of multiple components, or
of inhomogeneities, will naturally broaden the width.
As noted in \S2.1,
J1939$-$6342 (1934$-$638) has the narrowest width of the
previously identified GPS sources, with a value similar to
that of 0108+388. 
J1939$-$6342 has a compact double (CD)
morphology, 
and although 0108+388 was initially considered a CD, it is
now recognised as a CSO \citep{con94,tay96}.
Comparison with studies of other GPS sources
in the literature reveals that this appears to be a characteristic
of CD/CSO sources.
For example, the other two CD sources in the sample of 15 GPS sources
of \cite{ode91}, 0019$-$000 and 1225+368, 
have spectral FWHMs of 1$\sim$1.25 decades of
frequency, at the low end of the distribution.

To quantify this further, we have examined the sample of fainter GPS
sources of \cite{sne00}, which is presented with
spectra compiled from total flux density measurements and morphologies
determined from VLBI observations. (The criteria for morphological
classification are given in \cite{sne00}: we note
here that the CD classification required similar spectra in both components
and not necessarily similar flux densities.)
In Figure~\ref{Fig6}, the distributions of spectral widths are plotted for the 
14 CD/CSO sources and the 30 other sources from \cite{sne00}.
A Kolmogorov-Smirnov (KS)
test comparing two distributions yields a probability that
the two are drawn from the same distribution of 0.6\%.
A more conservative estimate, obtained by excluding the
narrowest CD/CSO width and the largest `other morphologies' width,
is 1.4\%. 
Such a trend is plausible, as CD/CSO sources are believed to
lie close to the plane of the sky and/or have generally small
advance speeds, resulting in emission on both sides of the core
being visible.
In contrast, core-jet sources have spectra
in which a flatter spectrum core component
makes an appreciable contribution, resulting in a broader peak.
In addition to the intrinsically narrower spectra, it may be
that CD and CSO sources may also have a sharper low frequency
cut-offs due to increased free-free absorption 
and/or synchrotron self-absorption.
There is, however, appreciable overlap between the two
distributions, and so the spectral width does not have a 
great predictive power in individual cases as to the morphology.
It will, however, be of interest to determine the morphology of
J1726$-$6427, the newly identified GPS with a narrow spectrum
(Fig.~\ref{Fig2}).

We have also compared the distribution of widths of all 28 sources
in Tables~\ref{tab1} and \ref{tab2} against all 44 sources in the 
\cite{sne00} sample:
the two distributions have a probability of 88\% of being drawn
from the same distribution.

\subsection{Redshift distribution}

As mentioned in \S1, a significant fraction of sources with GHz peaked
spectra have been found to lie at high-redshift.
This is borne out by the distribution of redshifts in Table~\ref{tab1}.
We can compare this with, e.g., the distribution of redshifts for 
277 sources in flat-spectrum sample of 
\cite{dri97}.
In that sample, 45\% of sources had $z<1$, 38\% had $1<z<2$,
13\% had $2<z<3$ and 4\% had $z>3$.
In contrast we find that 8 of the 17  (47\%) GPS sources
with known redshifts in Table~\ref{tab1} have $z>2$, and 4 of 17 (24\%) have
$z>3$. Of the newly identified GPS sources in Table~\ref{tab2}, two of five
(40\%) have $z>2$. The samples are small but the expected trend is
indeed observed.

The known and newly identified CSS sources and candidates 
show the opposite trend: five of seven sources (71\%) in Table~\ref{tab3}
and five of seven (71\%) in Table~\ref{tab4} have $z<1$. The marked difference
from the distribution of Drinkwater et al.\ is not unexpected, as
once again the CSS sources have steep spectra whereas the 
Drinkwater et al.\ sample was for flat-spectrum sources.
Of the 42 sources with known redshifts in the sample of CSS sources of 
\cite{fan90},
64\% had $z<1$ and 29\% had $1<z<2$, in broad agreement with
the distributions for the sample here.

In contrast, the distribution of redshifts for the sources in
Table~\ref{tab5} (neglecting the candidate CSO J1419$-$1928) has
8 of 13 (61\%) sources with $z<1$, four (31\%) with $1<z<2$, and one 
(8\%) with $z>2$ --- much closer to the distribution of
flat-spectrum quasars of \cite{dri97}. (We note that there is
some overlap between these samples, though the effect of this
is not significant.)

 \begin{figure*}
 \centering
 \includegraphics[angle=-90,width=12cm]{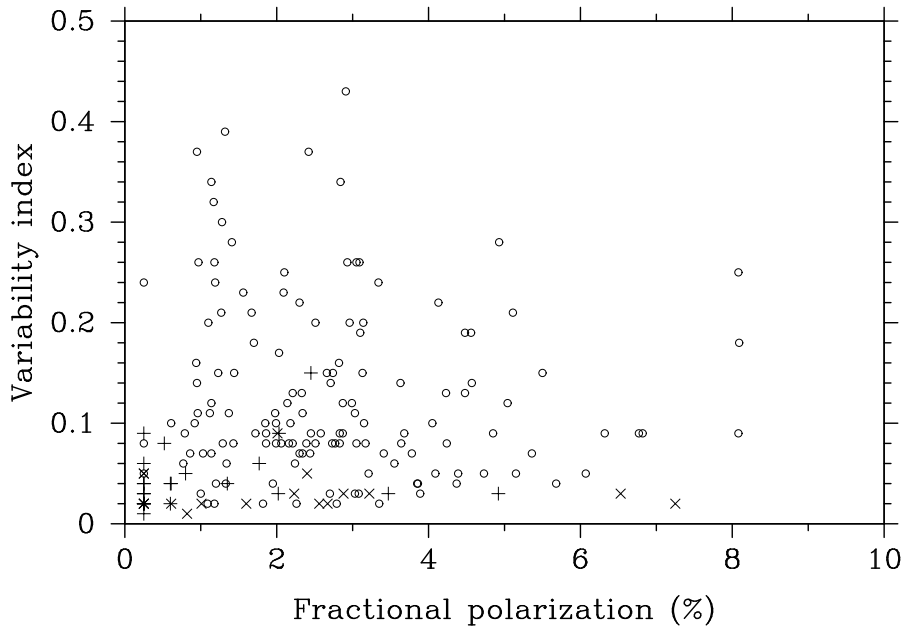}
 \caption{Variability index as a function of percentage linear polarization,
    for the 4.8\,GHz ATCA monitoring data \citep{tin03b}.
        Plus symbols are for the 28 GPS sources in Tables~\ref{tab1} 
    and \ref{tab2}, crosses for the
    18 CSS sources in Tables~\ref{tab3} and \ref{tab4} 
    for which these parameters were measured,
    and open circles for the 139 other sources.
    All the points with a plotted fractional polarization of 0.25
    represent upper limits of 0.5.
    A number of data points are over-plotted, particularly for GPS and CSS
    sources at low variability index and low polarization: there are, for example,
    7 GPS sources with an $m_{4.8}$ of 0.02 and $p_{4.8} <$0.5. 
 }
 \label{Fig7}%
 \end{figure*}

\subsection{Variability}

Comparing the 4.8\,GHz variability indices in Tables 
\ref{tab1}--\ref{tab4} with the
overall range and the median value (0.08) given in \S1
reveals that both the previously reported and newly identified
GPS and CSS sources have low variability indices.
This can also be seen in Figure~\ref{Fig7}, where
the 4.8\,GHz variability index has been plotted against
fractional polarization, for all 185 sources monitored in the ATCA program.
J1522$-$2730, previously reported
as a GPS source, has a relatively high variability
index and may, as noted by \cite{tor01} only show a
GPS-like spectrum at some epochs. 

More quantitatively, we can compare the distributions 
of variability indices for the three classes plotted in
Figure~\ref{Fig7}. The KS test confirms that the probability that the
distribution of the 28
GPS (previously reported and new candidate) sources
is drawn from the same distribution as for the 139 other 
(i.e., neither GPS nor CSS) sources is $<10^{-6}$.
The comparison of CSS (previously reported and new candidate) sources
and other sources yields a similar result. 
The probability that the distributions of GPS and CSS sources are
drawn from the same distribution is, in contrast, 19\%.

\cite{all02} have studied the long term variability of
a sample of 32 GPS sources and find that several sources exhibit
variability comparable to those seen in flat-spectrum, core-dominated
radio sources. Ten of 18 sources with good long-term monitoring data
show variability at some level, though the timescales are typically
longer than for flat-spectrum, core-dominated radio sources.  This is
also seen for a number of sources in this paper, which have low
variability indices over the 3.5~year period of the ATCA monitoring,
but which have clearly varied on longer timescales when comparing the
mean ATCA flux density with those from the PMN and PKS catalogs.

\cite{tor01} note, and we confirm, that the majority of their sources
with GPS-type spectra show higher variability than the established GPS
sources.  GPS-like spectra may be observed during certain stages
of large radio outbursts in flat-spectrum AGN, however the inverted
ATCA spectra in Figure~\ref{Fig5} appear, in comparison to the
admittedly small amount of supplementary data, to be relatively
persistent.  A number have higher turnover frequencies and/or quite
inverted spectra, and so these may be related to High Frequency
Peakers \citep{dal03}.

\subsection{Fractional polarization}

As was done for the variability indices,
we can compare the distributions 
of fractional polarization at 4.8\,GHz for the three classes plotted in
Figure~\ref{Fig7}. The probability that the
distribution of the GPS sources
is drawn from the same distribution as for the `other' 
sources is $<10^{-6}$.
For CSS sources, the probability is 3\%.
This is influenced by 
two CSS sources --- the previously reported J0627$-$0553,
and new candidate J0059$+$0006 ---
with fractional polarizations $>6$\%.
The probability that the distributions of GPS and CSS sources are
drawn from the same distribution is 3\%.
These results are in line with previous reports that CSS sources tend to have
higher fractional polarizations than GPS sources \citep[see, e.g.,][]{ode98}.

Nevertheless, sources with  low variability indices and
low fractional polarizations are much more likely to be
GPS or CSS sources.
Of the 24 sources with fractional polarizations
below 0.5\%, 21 appear in Tables \ref{tab1} to \ref{tab4}. 
Of the remaining three, one
is listed in Table~\ref{tab5}, and the other two are the extended radio galaxy
Cen~A and the gravitational lens system J1833$-$2103, both of which
are complex sources not well-characterised by the snapshot observing
mode used for the ATCA monitoring program \citep{tin03b}.

 \begin{figure*}
 \centering
 \includegraphics[angle=-90,width=18cm]{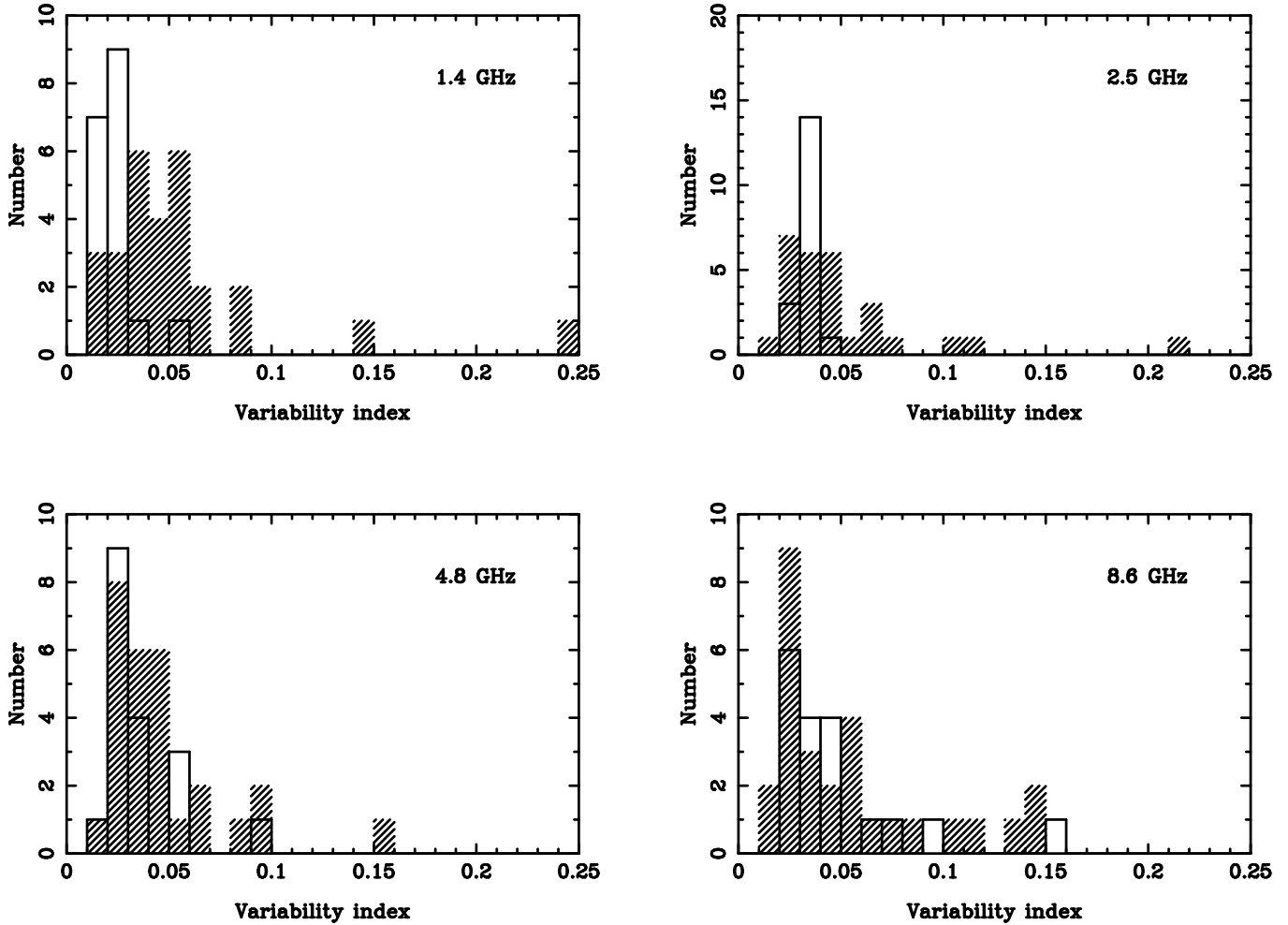}
 \caption{Distributions of variability index for GPS
    (hatched) and CSS (solid outline) sources.
    The distributions differ significantly at 1.4 and 2.5\,GHz,
    with CSS sources tending to be less variable.
 }
 \label{Fig8}%
 \end{figure*}

\subsection{Behaviour at other frequencies}

For the reasons outlined in \S\,1, we have focused on the
characteristics of sources at 4.8\,GHz. However, as the ATCA monitoring
program was undertaken at four frequencies, we are able to check
whether the conclusions drawn at 4.8\,GHz also apply at other
centimeter frequencies. We have performed the same KS tests 
as those described in the previous two sections
for variability and fractional polarization distributions 
at 1.4, 2.5 and 8.6\,GHz. 

At all frequencies, the distributions of
variability indices of both GPS and CSS sources
differ significantly (probabilities of being drawn from the
same distribution of $<$0.1\%) from those of the `other' sources.
At 1.4\,GHz the distributions of GPS and CSS sources are
significantly different, with the CSS sources being on average
less variable than the GPS sources.
At 2.5\,GHz the probability they
are drawn from the same distribution is 1\%, which increases
to 19\% at 4.8\,GHz and  69\% at 8.6\,GHz.
As shown in Figure~\ref{Fig8}, at 1.4 and 2.5\,GHz, the distributions of 
CSS sources are concentrated at lower variability indices.
The most variable GPS source at 1.4, 2.5 and 4.8\,GHz is 
J1522$-$2730 which, as discussed in \S\,2.1, is arguably not
a bona fide member of this class, however the significant differences
in distributions remain even if this source is excluded.
The lower variability of CSS sources at low frequencies is
also consistent with the more general trend (mentioned in \S6)
for increased variability as the spectral index increases.

The distributions of
fractional polarization of GPS sources
differ significantly at all frequencies
from those of `other' sources.
However, as frequency increases, 
the fractional polarization distributions of CSS sources become 
more similar to those of `other' sources, from being significantly
different at 1.4 and 2.5\,GHz, to a probability of 3\% at  4.8\,GHz,
and a probability of 31\% at 8.6\,GHz. The tendency for CSS
to become more polarized with increasing frequency (at least at centimeter
wavelengths) has been noted previously \citep[see, e.g.,][]{ode98}.
As a result, the distributions of GPS and CSS sources tend to 
become less similar with increasing frequency: the KS probabilities
are 51\% at 1.4\,GHz, 73\% at 2.5\,GHz, 3\% at 4.8\,GHz and 5\% at 8.6\,GHz.

It should be kept in mind that, although the new candidate GPS and
CSS sources have been selected primarily based on their spectra,
the variability indices and fractional polarizations have also been
considered. Thus the probabilities given above will, to some degree, reflect
this. The general trends that are observed, however, are unlikely to be
significantly affected.

\subsection{Low frequency turn-ups}

Early searches for GPS sources 
rejected candidates with a turn-up at low frequencies 
\citep[e.g.,][]{gop83}. However,
given the evidence that at least some GPS sources are
undergoing renewed activity, it is possible that
some GPS sources will have the GPS spectrum from the central kpc-scale
region imposed upon the extended steep spectrum component, resulting
in an overall spectrum that turns up at low frequencies.
In Figure~\ref{Fig1} it is apparent that J0241$-$0815 (NGC1052), J1522$-$2730,
and J2136+0041 show departures from the fitted spectral form.
Similar behaviour is seen for the new candidates J1624$-$6809
and, possibly, J1837$-$7108 (Fig.~\ref{Fig2}). 
In extreme cases, such as J0745$-$0044,
J0900$-$2808 (both Fig.~\ref{Fig1}) and J1658$-$0739 (Fig.~\ref{Fig2}),
the spectra clearly turn up at low frequencies.
However, these sources have 
clear spectral peaks at GHz frequencies, and display the 
variability and polarization properties of other GPS sources.
Imaging observations are required to determine whether the
low frequency component corresponds to extended emission within
which the GHz-peaked spectrum component is embedded.

 \begin{figure*}
 \centering
 \includegraphics[angle=-90,width=6cm]{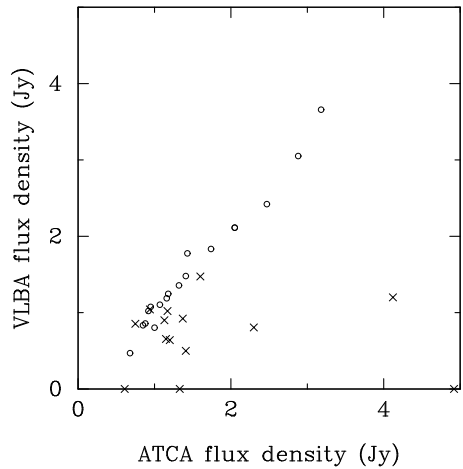}
 \caption{Comparison of mean 4.8\,GHz flux density on the ATCA 6\,km baseline
    \citep{tin03b} with the total model-fit flux density from 
    the 4.8\,GHz VLBA pre-launch survey
    observations in June 1996 \citep{fom00}. 
    GPS sources and candidates are plotted with circles, and
    CSS sources and candidates are plotted with crosses. For clarity, one GPS
    source has been excluded from this plot: J2136+0041, which has an ATCA
    flux density of 9.47\,Jy and a model-fit VLBA flux density of
    10.13\,Jy. The expected trend is observed, with GPS sources being more
    compact (i.e. having most of the flux density recovered on VLBI
    baselines) than CSS sources.  
 }
 \label{Fig9}%
 \end{figure*}

\subsection{Compactness}

In the ATCA monitoring program, as described in \S\,1, 
a flag was assigned according to the degree of 
compactness of a source on the maximum 6\,km baseline
\citep[see also][]{tin03b}.
GPS sources are expected to be confined to sub-kpc scales,
with CSS sources generally $<$20~kpc in extent.
This trend is seen in the distribution of flags for
Tables \ref{tab1} and \ref{tab2} (18~`c', 3 `e', 7 `l')
as compared to those of Table \ref{tab3} and \ref{tab4} (9 `c', 9 `e', 3 `l').

We can investigate this more quantitatively by comparing the mean ATCA
4.8\,GHz correlated flux density on the 6\,km baseline with the
sum of the model-fit components from the 4.8\,GHz VLBA observations of
June 1996 \citep{fom00}.  The results are shown in Figure~\ref{Fig9}.  The
expected trend is seen, with most of the flux density of GPS sources
being recovered in the VLBI observations, whereas the CSS sources were
generally partially (and in some cases completely) resolved.

\section{Conclusions}

The multi-frequency, multi-epoch monitoring of 185 radio sources 
(and multi-frequency data of 17 more) with the
Australia Telescope Compact Array has provided a data set 
well-suited to searching for new GPS and CSS candidates.
We have studied the properties of previously reported GPS and CSS
sources, and identified eight new candidate GPS sources
and 12 new candidate CSS sources with similar properties.
VLBI studies are required for 
a number of these candidates to determine whether they satisfy the compactness 
requirements for membership of their respective classes.
The incidence of GPS sources in the ATCA sample is in agreement
with that of other samples, with the incidence of CSS sources 
being lower than that of other samples due to the preference
given to flatter spectral index sources in selecting the ATCA sample.
We have confirmed that GPS sources are generally compact on 6\,km
baselines, and that in most cases the ATCA flux densities are
in good agreement with the flux densities recovered in VLBI 
imaging, whereas CSS sources are in general less compact and are
sometimes completely resolved in VLBI observations.
The ATCA monitoring at 4.8\,GHz confirms the expectation that
both GPS and CSS sources have low variability indices 
and low fractional polarizations.
These conclusions hold true for the other frequencies, though
with the fractional polarization of CSS sources seen to increase
with frequency, as has been previously noted. We also find that
CSS sources are on average less variable at 1.6 and 2.5\,GHz
than GPS sources.
Based on qualitative observations of sources in the ATCA sample,
and a quantitative study of the faint GPS sample of \cite{sne00},
we have also found that GPS sources with Compact Double and
Compact Symmetric Object morphologies tend to have narrower
spectra than other classes of GPS sources, compatible with 
a higher degree of homogeneity in these sources, a larger viewing angle
to the jet axis, and/or increased absorption at lower frequencies.

\begin{acknowledgements}
This research has made use of NASA's Astrophysics Data System
Bibliographic Services, and the NASA/IPAC Extragalactic Database (NED)
which is operated by the Jet Propulsion Laboratory, California
Institute of Technology, under contract with the National Aeronautics
and Space Administration. Ignas Snellen is thanked for helpful
discussions, and the referee is acknowledged for comments and suggestions
which have resulted in an improved paper.
\end{acknowledgements}

\end{document}